\pdfoutput=1
\documentclass[5p,twocolumn]{elsarticle}
\usepackage{graphicx,epstopdf,titlesec,siunitx,amsmath,fixltx2e,subcaption,textgreek,amsmath,lineno,tabularx,makecell,xcolor,colortbl}
\journal{Icarus}
\biboptions{authoryear}
\newcolumntype{L}{>{\raggedright\arraybackslash}X}

\begin{document}

\definecolor{grey}{rgb}{0.9,0.9,0.9}

\begin{frontmatter}

\title{Vertically-resolved observations of Jupiter's quasi-quadrennial oscillation from 2012 to 2019}

\author[swri]{R. S. Giles\corref{cor1}}
\ead{rgiles@swri.edu}
\author[swri]{T. K. Greathouse}
\author[gsfc,cresst]{R. G. Cosentino}
\author[jpl]{G. S. Orton}
\author[austin]{J. H. Lacy}

\address[swri]{Southwest Research Institute, San Antonio, TX, USA}
\address[gsfc]{ NASA Goddard Spaceflight Center, Greenbelt, MD, USA}
\address[cresst]{CRESST II and Department of Astronomy, University of Maryland, College Park, MD, USA}
\address[jpl]{Jet Propulsion Laboratory / California Institute of Technology, Pasadena, CA, USA}
\address[austin]{University of Texas at Austin, Austin, TX, USA}

\cortext[cor1]{Corresponding author}

\begin{abstract}

Over the last eight years, a rich dataset of mid-infrared CH\textsubscript{4} observations from the TEXES instrument at IRTF has been used to characterize the thermal evolution of Jupiter's stratosphere. These data were used to produce vertically-resolved temperature maps between latitudes of 50$^{\circ}$S and 50$^{\circ}$N, allowing us to track approximately two periods of Jupiter's quasi-quadrennial oscillation (QQO). During the first five years of observations, the QQO has a smooth sinusoidal pattern with a period of 4.0$\pm$0.2 years and an amplitude of 7$\pm$1 K at 13.5 mbar (our region of maximum sensitivity). In 2017, we note an abrupt change to this pattern, with the phase being shifted backwards by $\sim$1 year. Searching for possible causes of this QQO delay, we investigated the TEXES zonally-resolved temperature retrievals and found that in May/June 2017, there was an unusually warm thermal anomaly located at a latitude of 28$^{\circ}$N and a pressure of 1.2 mbar, moving westward with a velocity of 19$\pm$4 ms\textsuperscript{-1}. We suggest that there may be a link between these two events.  

\end{abstract}


\end{frontmatter}



\section{Introduction}

Long-term observations of Jupiter have shown periodic variations in the planet's stratospheric temperatures. Between 1980 and 2001, 7.8-\textmu m images of Jupiter from NASA's Infrared Telescope Facility (IRTF) were used to track the brightness temperature of Jupiter's atmosphere in the 10--20 mbar region~\citep{orton91,leovy91,friedson99,simon-miller06}. These observations showed an oscillation in the equatorial and low-latitude regions of the planet with a period of $\sim$4 years; the stratospheric temperatures oscillate between having a local maximum at the equator and local minima at planetocentric latitudes of $\pm$13$^{\circ}$, and having a local minimum at the equator and local maxima at planetocentric latitudes of $\pm$13$^{\circ}$. This phenomenon was named the quasi-quadrennial oscillation (QQO) by~\citet{leovy91}, who noted its resemblance to the quasi-biennial oscillation (QBO) that is seen in the Earth's stratosphere. 

The QBO is a quasi-periodic oscillation in the Earth's stratospheric zonal mean winds at tropical latitudes~\citep{baldwin01}. At a given pressure level, the equatorial wind direction oscillates between easterly and westerly with a mean period of 28 months. The easterly and westerly wind regimes propagate downwards with time, so there is a phase offset between different pressure levels and a vertical wind shear that alternates between positive and negative. On both Earth and Jupiter, the mean zonal flow can be assumed to be in approximately hydrostatic and geostropic balance~\citep{leovy91}, which means that the vertical wind shear and the horizontal temperature gradient are coupled through the thermal wind equation~\citep{baldwin01}. Unlike on the Earth, it is difficult to make direct wind measurements in Jupiter's stratosphere, but it is possible to study the QQO's signature and evolution by observing changes in the temperature field.

Despite being an equatorial phenomenon, the QBO has a significant impact on the dynamics of the Earth's atmosphere at all latitudes; this includes affecting the strength of Atlantic hurricanes and the breakdown of the wintertime stratospheric polar vortices~\citep{baldwin01}. The QQO may also be an important factor in Jupiter's weather patterns, and close monitoring of the QQO phase pattern is required to make any correlations with observed dynamical changes. Careful analysis of the QQO also provides insight into the waves that are thought to drive the phenomenon. The Earth's QBO is driven by the absorption of vertically-propagating atmospheric waves that break and deposit their zonal momentum at a critical level that is determined by the background zonal flow~\citep{baldwin01}. For the Earth, \citet{dunkerton97} showed that a combination of Kelvin, Rossby-gravity, inertia-gravity, and smaller-scale gravity waves are required to reproduce observed winds and temperatures. Similar mechanisms have been proposed as the driving forces behind Jupiter's QQO, and observations play an important role in constraining atmospheric models.

The first models of the QQO were developed by \citet{friedson99} and \citet{li00}, who were both able to reproduce QQO-like phenomena within mechanistic models of Jupiter's atmosphere; \citet{li00} suggested that equatorial oscillations could be produced using a combination of planetary-scale waves, while \citet{friedson99} implemented a flat-spectrum gravity wave drag parameterization and compared the effects of gravity waves and planetary-scale waves, finding that small-scale, short-period gravity were better able to drive a QQO-like phenomenon. However, these studies were limited by the observations that were available at the time. The 7.8-\textmu m broad-band imaging campaign provided a valuable dataset over a long timeframe, but was limited to a single pressure range and did not have any sensitivity to vertical variations in the thermal profile.

More recent studies of Jupiter's QQO have made use of spacecraft data from Voyager IRIS~\citep{simon-miller06} and Cassini CIRS~\citep{flasar04,simon-miller06}. Both of these instruments provide spectroscopic observations of Jupiter, allowing vertical temperature profiles to be retrieved in the stratosphere. This vertical information is an important addition to the studies of the QQO, but as the spacecraft observations were obtained during flybys of Jupiter, they were each restricted to short timeframes and cannot be used to study how the QQO varies with time. 

In order to improve our three-dimensional understanding of the QQO and to inform future modeling efforts, we began a long-term study of Jupiter's stratospheric temperatures in 2012, using the Texas Echelon Cross Echelle Spectrograph~\citep[TEXES,][]{lacy02}, which operates as a visitor instrument at the IRTF. TEXES provides high-resolution (R=80,0000) spectroscopy of line profiles, which allows us to vertically resolve the temperature structure in Jupiter's stratosphere to a greater degree than was possible with the lower resolution Voyager IRIS and Cassini CIRS observations. After the first complete cycle was observed (2012--2016), the data were used to constrain a new General Circulation Model (GCM) analysis by~\citet{cosentino17}. The TEXES observations showed that the QQO extends upwards to lower pressures ($\sim$2 mbar) than previously known, and that there is a smooth sinusoidal transition between maxima and minima over a large range of pressures. \citet{cosentino17} concluded that high-frequency gravity waves are significant contributors of momentum to the QQO, agreeing with earlier results from \citet{friedson99}. In particular, \citet{cosentino17} found that a stochastic gravity wave drag parameterization, representative of an atmosphere with convection, is better able to model the observations.

We have now been observing Jupiter's QQO with TEXES/IRTF for approximately eight years, which corresponds to two cycles. These data allow us to look for additional variability superimposed on top of the QQO. In 2015--16, an unusual disruption to the Earth's QBO was noted by~\citet{osprey16} and~\citet{newman16}. The exact causes of this event are still unclear, but both studies suggest it may be due to intrusion of Rossby waves from the winter hemisphere into the equatorial region of the planet. Like Jupiter, Saturn also has an oscillation in the equatorial stratosphere, known as Saturn's Semi-Annual Oscillation (SSAO), and this has been observed to be disrupted by a large long-lasting storm in the northern mid-latitudes~\citep{fletcher17b}.

In this paper, we present an updated vertically resolved time series of Jupiter's QQO covering the 2012--2019 time period. We find that after five years of smoothly varying sinusoidal behaviour, there was a change in the QQO behaviour in 2017, and we use our spatially resolved observations to show that this occurred in conjunction with a localized stratospheric thermal anomaly. The TEXES observations and the data reduction process are described in Section~\ref{sec:observations} and the radiative transfer model is presented in Section~\ref{sec:modeling}. The long-term QQO observations and the 2017 disruption are discussed in Section~\ref{sec:results} and the conclusions are summarized in Section~\ref{sec:conclusions}.


\section{Observations}
\label{sec:observations}

\subsection{TEXES observations}

Between 2012 and 2019, high-resolution mid-infrared observations of Jupiter were made with the TEXES instrument, mounted at the 3-m NASA Infrared Telescope Facility. Data were obtained during eighteen observing runs, summarized in Table~\ref{tab:run_summary}. TEXES is a cross-dispered grating spectrograph that covers wavelengths of 4.5--25 \textmu m in the mid-infrared, with a resolving power between 4,000 and 80,000 depending on the mode of operation. The scan-map mode of TEXES produces spectral datacubes (two spatial dimensions and one spectral dimension) of extended objects by stepping the slit across the sky perpendicular to the slit length.

In order to study Jupiter's stratospheric temperatures, we used observations centered at 1247.5 cm\textsuperscript{-1} (8.02 \textmu m), with a spectral bandpass of $\sim$7 cm\textsuperscript{-1}. This spectral range covers six strong P-branch lines of the CH\textsubscript{4} \textnu\textsubscript{4} band, which is centered at 1306 cm\textsuperscript{-1}~\citep{brown03}. TEXES was used in its highest spectral resolution mode (R=80,000), which means that these lines are spectrally resolved. CH\textsubscript{4} is homogeneously mixed in Jupiter's stratosphere, and its abundance is well known. Variations in the shapes of CH\textsubscript{4} emission lines are due to temperature variations alone and these emission lines therefore provide a direct measurement of Jupiter's stratospheric temperature. At the center of the strong emission lines, the atmosphere becomes optically thick at high altitudes in the stratosphere, so we are sensitive to temperatures at those high altitudes. In the wings of those strong lines, or in the center of weaker lines, the atmosphere does not become opaque until a lower altitude, and so we can probe deeper into the stratosphere. By simultaneously modeling the 1244--1251 cm\textsuperscript{-1} spectral range, we are able to retrieve the vertical temperature profile in the 0.1--30 mbar pressure region.

Three-dimensional temperature maps covering the 50$^{\circ}$S--50$^{\circ}$N latitude range were built up by using the TEXES scan-map mode. We oriented the slit along the celestial north-south direction and scanned across the planet from west to east. We used a step size of 0.7$^{\prime\prime}$ (half the width of the slit) and observations of sky were included at the beginning and end of each scan to aid with sky subtraction. In the high-resolution observing mode, the TEXES slit length is 6$^{\prime\prime}$, which is considerably smaller than the diameter of Jupiter (see Table~\ref{tab:run_summary}). In order to cover the entire 50$^{\circ}$S--50$^{\circ}$N latitude range, we therefore performed multiple scans with different north-south offsets. This observing sequence was repeated continuously for $\sim$6 hours on a single night; due to Jupiter's rotation period of 10 hours, this provides coverage of over half of the planet. When possible, observations on a subsequent night were then used to complete the global map. Table~\ref{tab:run_summary} shows that we were able to obtain full longitudinal coverage at the equator on the vast majority of the observing runs.

At 1247.5 cm\textsuperscript{-1}, the diffraction limit at the 3-m IRTF is 0.7$^{\prime\prime}$, which is comparable to the typical seeing at the telescope. For the angular diameters provided in Table~\ref{tab:run_summary}, this is equivalent to a spatial resolution of $\sim$2$^{\circ}$ latitude at the equator. In the longitudinal direction, the 1.4$^{\prime\prime}$ slit width leads to a spatial resolution of $\sim$4$^{\circ}$ longitude.

\begin{table*}
\begin{center}
\begin{tabular}{|>{\centering\arraybackslash}m{1.8cm} | >{\centering\arraybackslash}m{2.4cm} | >{\centering\arraybackslash}m{1.6cm} | >{\centering\arraybackslash}m{2.2cm} | >{\centering\arraybackslash}m{2.2cm} |
>{\centering\arraybackslash}m{1.7cm} |} 

 \hline

Date & Longitudinal coverage at the equator (\%) & Angular diameter (arcsec) & Earth-Jupiter velocity (km/s) & Sub-solar planetocentric latitude ($^{\circ}$) & Heliocentric distance (AU) \\

 \hline

Jan 2012 & 72 & 41 &  28 &  3.07 & 4.98 \\
\rowcolor{grey} Sep 2012 & 100 & 44 & -24 &  2.92 & 5.04 \\
Feb 2013 & 100 & 42 &  27 &  2.68 & 5.08 \\
\rowcolor{grey} Feb 2014 & 100 & 43 &  23 &  1.44 & 5.21 \\
Dec 2014 & 78 & 41 & -25 &  0.23 & 5.31 \\
\rowcolor{grey} Mar 2015 & 100 & 41 &  25 & -0.26 & 5.35 \\
Nov 2015 & 100 & 34 & -26 & -1.17 & 5.41 \\
\rowcolor{grey} Jan 2016 & 100 & 41 & -24 & -1.39 & 5.42 \\
Apr 2016 & 100 & 41 &  24 & -1.76 & 5.44 \\
\rowcolor{grey} Jan 2017 & 100 & 37 & -28 & -2.51 & 5.46 \\
May 2017 & 100 & 41 &  23 & -2.79 & 5.45 \\
\rowcolor{grey} Jul 2017 & 61 & 36 &  27 & -2.86 & 5.45 \\
Feb 2018 & 100 & 37 & -29 & -3.07 & 5.43 \\
\rowcolor{grey} Jul 2018 & 100 & 40 &  25 & -3.07 & 5.40 \\
Sep 2018 & 100 & 34 &  22 & -3.02 & 5.38 \\
\rowcolor{grey} Feb 2019 & 100 & 35 & -26 & -2.84 & 5.34 \\
Apr 2019 & 100& 42 & -23 & -2.72 & 5.32 \\
\rowcolor{grey} Aug 2019 & 100 & 40 &  25 & -2.44 & 5.28 \\

 \hline

\end{tabular}
\caption{Summary of IRTF observing runs contributing to this study: the month and year, the longitudinal coverage obtained over the course of the observing run, Jupiter's angular diameter as observed from the Earth, the Doppler velocity, the sub-solar Jovian latitude and the distance of Jupiter from the sun.}
\label{tab:run_summary}
\end{center}
\end{table*}

Our long-term observing program was designed to study both long- and short-term variability in Jupiter's stratospheric temperatures. In order to track the long-term trends in temperature, observations were made on several observing runs per year. These runs were nominally planned as pairs with a 1-month spacing about each quadrature. Observing near quadrature ensured a sufficiently large Doppler shift between Earth and Jupiter to separate the terrestrial and Jovian CH\textsubscript{4} lines (see Table~\ref{tab:run_summary} for the Doppler velocity during each run). In order to measure short-term variability, we aimed to repeat our observations at both the beginning and end of a given observing run, however weather conditions meant that this was not always possible. In total, these observations provide comparisons over 1-week, 1-month and 4-month intervals.

\subsection{Data reduction}

The individual scans of Jupiter were each reduced using the data reduction pipeline described in~\citet{lacy02}. This software performs flat fielding and sky subtraction, and then removes instrumental geometric optical distortions. Wavelength calibration was achieved by using telluric absorption lines and radiometric calibration was achieved by using the measured radiance of a room temperature blackbody that is automatically placed in front of instrument aperture prior to each set of observations. The noise of each spectral data point was calculated following the method described for TEXES observations in~\citet{greathouse11}.

The pipeline-reduced data products were then geometrically calibrated, cylindrically projected and co-added. The scan parameters were used to calculate the shape of the planet's limb, which was then fitted to each observation by eye. This allows each pixel to be assigned a latitude and longitude. The Doppler shift of the spectrum was calculated for each pixel, comprising both a component due to the Earth-Jupiter velocity and a position-dependent component due to Jupiter's rotation (zonal wind speeds are negligible in comparison); this Doppler shift was removed for each pixel, shifting the Jovian spectrum into the rest frame. The observations were then binned according to airmass and then projected onto a latitude-longitude grid; this allows multiple scans to be co-added to produce a single data cube per airmass bin.


\section{Spectral modeling}
\label{sec:modeling}

The TEXES spectra were modeled using a radiative transfer and retrieval code that has previously been used to model the stratospheres of Saturn~\citep{greathouse05} and Neptune~\citep{greathouse11}. The code is made up of a line-by-line radiative transfer code that calculates the theoretical spectrum for a given atmospheric profile and an optimal estimation retrieval code that uses a Levenberg-Marquardt approach to iteratively adjust the atmospheric parameters and achieve an optimal fit between the theoretical spectrum and the observed spectrum~\citep{rodgers00}. The code is described in greater depth in~\citet{greathouse11}.

For this study, Jupiter's atmosphere was divided into 95 levels between 4 bar and $1\times10^{-7}$ bar, equally spaced in log(p). The vertical distribution of CH\textsubscript{4} was obtained from the photochemical model of~\citet{moses05}, which takes into account both eddy mixing and molecular diffusion. The line data for \textsuperscript{12}CH\textsubscript{4}, \textsuperscript{13}CH\textsubscript{4} and \textsuperscript{12}CH\textsubscript{3}D were obtained from the HITRAN molecular database~\citep{rothman05}, and the temperature dependence of the line widths in an H\textsubscript{2} atmosphere was obtained from~\citet{margolis93}. Collision induced absorption data for H\textsubscript{2}-H\textsubscript{2}, H\textsubscript{2}-He and H\textsubscript{2}-CH\textsubscript{4} were included from~\citet{orton07},~\citet{borysow88} and~\citet{borysow86} respectively. We assume local thermodynamic equilibrium (LTE) at all pressure levels; our observations are primarily sensitive to pressures in the 0.1--30 mbar range, and the transition between LTE and non-LTE does not occur until $\sim$1 \textmu bar~\citep{drossart99}.

\begin{figure*}
\centering
\includegraphics[width=12cm]{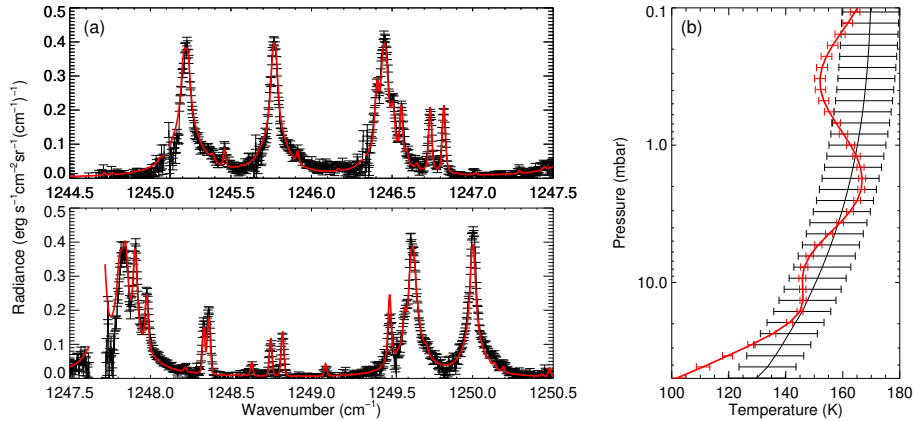}
\caption{(a) An example of a zonally-averaged equatorial TEXES spectrum from 17--18 November 2015 (black) along with the best-fit model spectrum (red). The spectrum is split into two frames to make the features more clear. (b) Corresponding vertical temperature profile (red), along with the a priori temperature profile (black).}
\label{fig:spectrum}
\end{figure*}

The a priori vertical temperature profile is based on the model described in~\citet{moses05}. This profile was allowed to vary in the retrievals, in order to fit the TEXES spectra. The temperature is retrieved at each of the 95 pressure levels and a correlation length of 1 scale height is used to prevent unphysically sharp deviations between adjacent levels. Figure~\ref{fig:spectrum} shows an example of a fitted spectrum and a retrieved temperature profile for data obtained on 17--18 November 2015. Figure~\ref{fig:spectrum}(a) shows the TEXES spectrum in black, along with the best-fit model spectrum in red. The gap in the data at 1247.7 cm\textsuperscript{-1} is due to the presence of a strong telluric line. Figure~\ref{fig:spectrum}(b) shows the corresponding retrieved vertical temperature profile along with the a priori temperature profile. The error bars shown in Figure~\ref{fig:spectrum}(b) represent the formal uncertainty in the retrieval. They are a minimum at $\sim$13 mbar, the pressure level of maximum sensitivity. As the pressure increases, the formal error bars also increase, but do not fully capture the dependence on any assumed model parameters in the deeper atmosphere. We consider the retrievals to be robust in the 0.1--30 mbar region. Within this region, the formal error bars are $\sim$1.5 K; in the subsequent sections, we adopt a more conservative error of 2 K to include systematic effects~\citep{cosentino17}.  

The radiative transfer and retrieval code was applied to all of the observations described in Section~\ref{sec:observations} in order to produce both two-dimensional and three-dimensional temperature maps for each observing date. As the QQO is an oscillation pattern in the global-scale wind field, we are primarily interested in the two-dimensional, zonally-averaged temperature maps. As described in Section~\ref{sec:observations}, the TEXES spectral data were binned according to latitude, longitude and Jovian airmass. Data obtained from several nights on each observing run were then co-added to provide as complete longitudinal coverage as possible. The data were then averaged across all longitudes, in order to produce a single spectrum for each latitude (2$^{\circ}$ bins) and airmass (four bins, covering airmasses of 1--5). For each latitude, the retrieval code was used to simultaneously fit the different airmass spectra, in order to produce a final vertical temperature profile for that latitude. Combining the vertical temperature profiles for each latitude produces a two-dimensional temperature map. These zonally-averaged maps are the focus of Sections~\ref{sec:longterm} and~\ref{sec:disruption}.

The observational data were also used to produce zonally-resolved three-dimensional maps. These were generated by retrieving the vertical temperature profile at each longitude and latitude point, instead of averaging across the longitudinal dimension. This allows us to search for spatially-localized thermal anomalies, as shown in Section~\ref{sec:anomaly}. 

\begin{figure*}
\centering
\includegraphics[width=12cm]{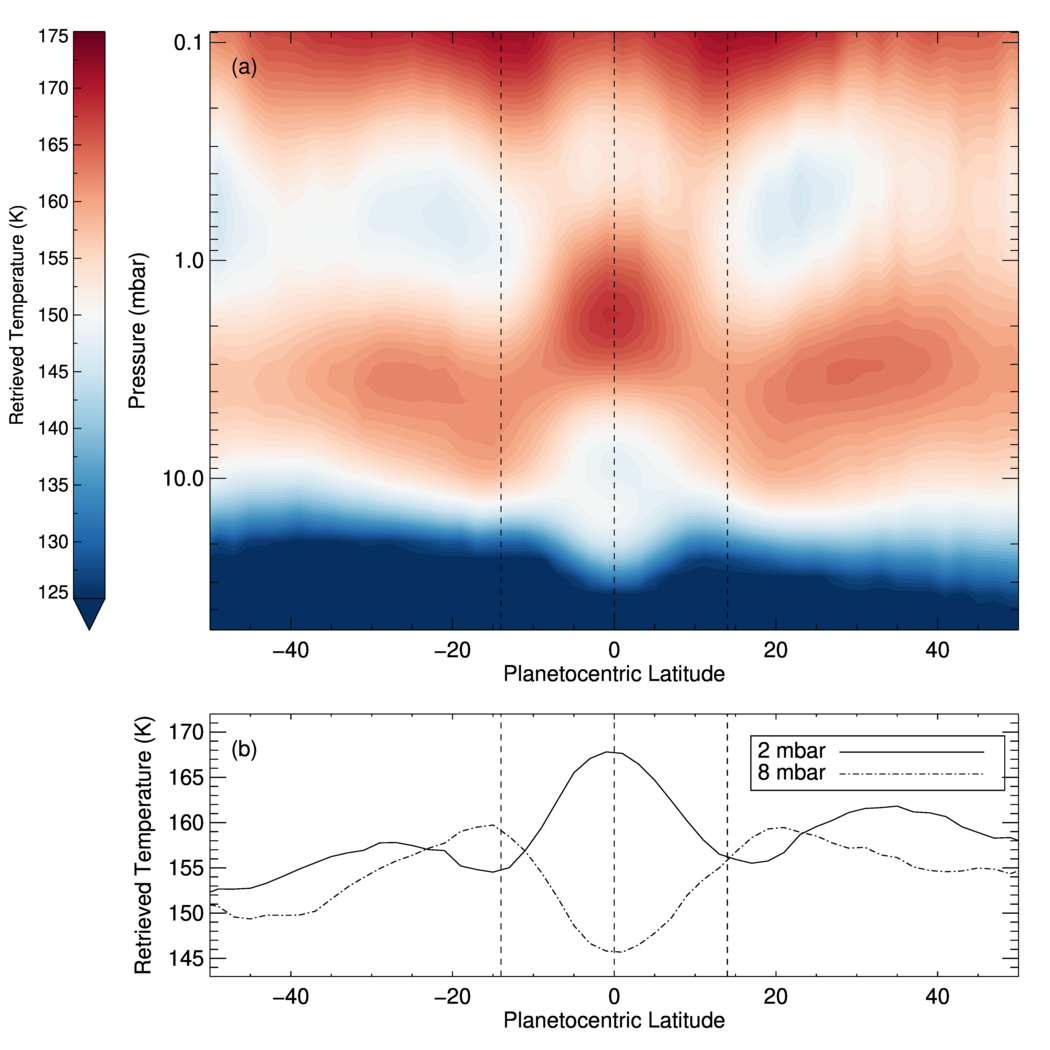}
\caption{(a) Example of a zonally-averaged retrieved temperature map from 17--18 November 2015. The vertical dashed lines show latitudes of 14$^{\circ}$S, 0$^{\circ}$ and 14$^{\circ}$N, the three latitudes that are shown in Figure~\ref{fig:time_series}. (b) Slices through (a) at two different pressures levels representing a local maximum at the equator (2 mbar) and a local minimum at the equator (8 mbar).}
\label{fig:zonally_averaged}
\end{figure*}

An example of a retrieved zonally-averaged temperature map from 17--18 November 2015 is shown in Figure~\ref{fig:zonally_averaged}(a). The dashed lines in the figure show latitudes of 14$^{\circ}$S, 0$^{\circ}$ and 14$^{\circ}$N, the locations of the main equatorial QQO signature and the anti-phase components. Figure~\ref{fig:zonally_averaged}(a) shows a clear vertical oscillation in temperature at the equator, with a strong local minimum (relative to the off-equatorial $\pm$14$^{\circ}$ temperatures) at approximately 8 mbar and strong local maxima at approximately 2 mbar and 20 mbar. This is highlighted by Figure~\ref{fig:zonally_averaged}(b), which shows the temperatures as a function of latitude for 2 mbar and 8 mbar.

This vertical oscillation is a signature of the QQO; as time passes the local maxima and minima at the equator move to deeper pressure levels. To study this, comparable temperature maps were made for each observing run conducted in 2012--2019 (Table~\ref{tab:run_summary}). The evolution of the stratospheric temperatures with time is described and discussed in Section~\ref{sec:results}. Section~\ref{sec:longterm} discusses the evolution of Jupiter's QQO over the 2012-2017, while Section~\ref{sec:disruption} discusses an apparent disruption in the QQO pattern in 2017.

We note that Figure~\ref{fig:zonally_averaged} also shows other interesting phenomena, such as a vertical temperature oscillation in the mid-latitudes ($\pm$20--30$^{\circ}$). However, no oscillation is observed at these latitudes and there is therefore no clear link with the QQO. Analysis of the retrieved mid-latitude temperatures is therefore outside the scope of this paper.

\section{Results and discussion}
\label{sec:results}

\subsection{Behavior of the QQO in 2012--2017}
\label{sec:longterm}

\begin{figure*}
\centering
\includegraphics[width=12.5cm]{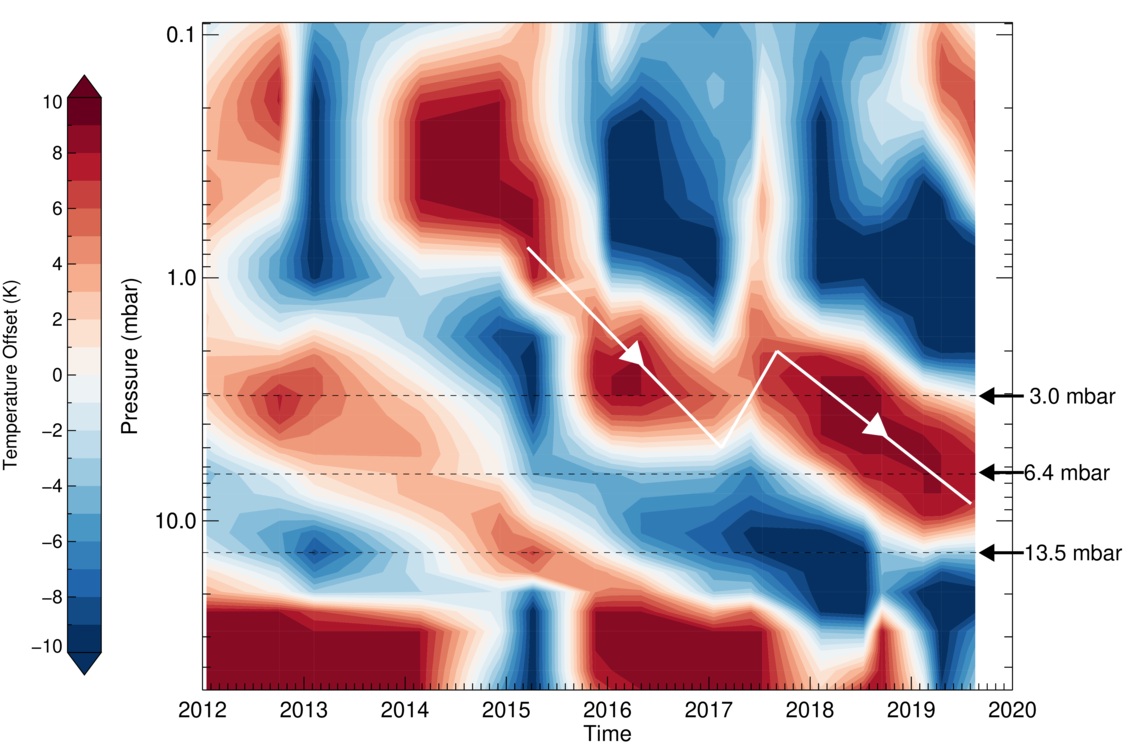}
\caption{Time series of Jupiter's QQO from 2012--2019 as a function of pressure and time. The retrieved temperatures are from Jupiter's equator (2$^{\circ}$S--2$^{\circ}$N) and the temperature offset for each pressure level is defined as the zonally-averaged temperature relative to the average of the maximum and minimum temperature during the first four years of data (corresponding to a complete, unperturbed oscillation). The dashed black lines highlight the 3.0-mbar, 6.4-mbar and 13.5-mbar pressure levels that are shown in Figure~\ref{fig:time_series}. The white line shows an apparent disruption in the descending warm branch.}
\label{fig:time_series_contour}
\end{figure*}

Figure~\ref{fig:time_series_contour} shows the time series of Jupiter's equatorial temperature oscillation during the entire 2012--2019 observation period. All observations from a given observing run are combined into a single zonally-averaged temperature retrieval. The retrieved temperatures are the average from the 2$^{\circ}$S--2$^{\circ}$N latitude range and the temperature offset at each pressure level is defined as the zonally-averaged temperature relative to the average of the maximum and minimum temperature in 2012--2016 (corresponding to a complete, unperturbed oscillation).

Figure~\ref{fig:time_series_contour} is analogous to the time-height diagrams of the Earth's QBO, as seen in e.g.~\citet{naujokat86}. As with the QBO, the QQO consists of downward propagating warm and cool regions, causing the temperature at a given pressure level to oscillate with time. With the Earth's QBO, the time taken for the pattern to repeat is approximately two years, and for Jupiter's QQO, the period is approximately four years. On Earth, wind speeds can be measured directly and so time-height diagrams are typically presented in terms of winds. However, the wind shear and temperatures are coupled via the thermal wind equation ~\citep[which for Earth remains valid deep into the tropics,][]{allen07}; positive (eastward/westerly) vertical wind shear in the QBO is correlated with warm equatorial anomalies~\citep{pascoe05}.

\begin{figure*}
\centering
\includegraphics[width=12cm]{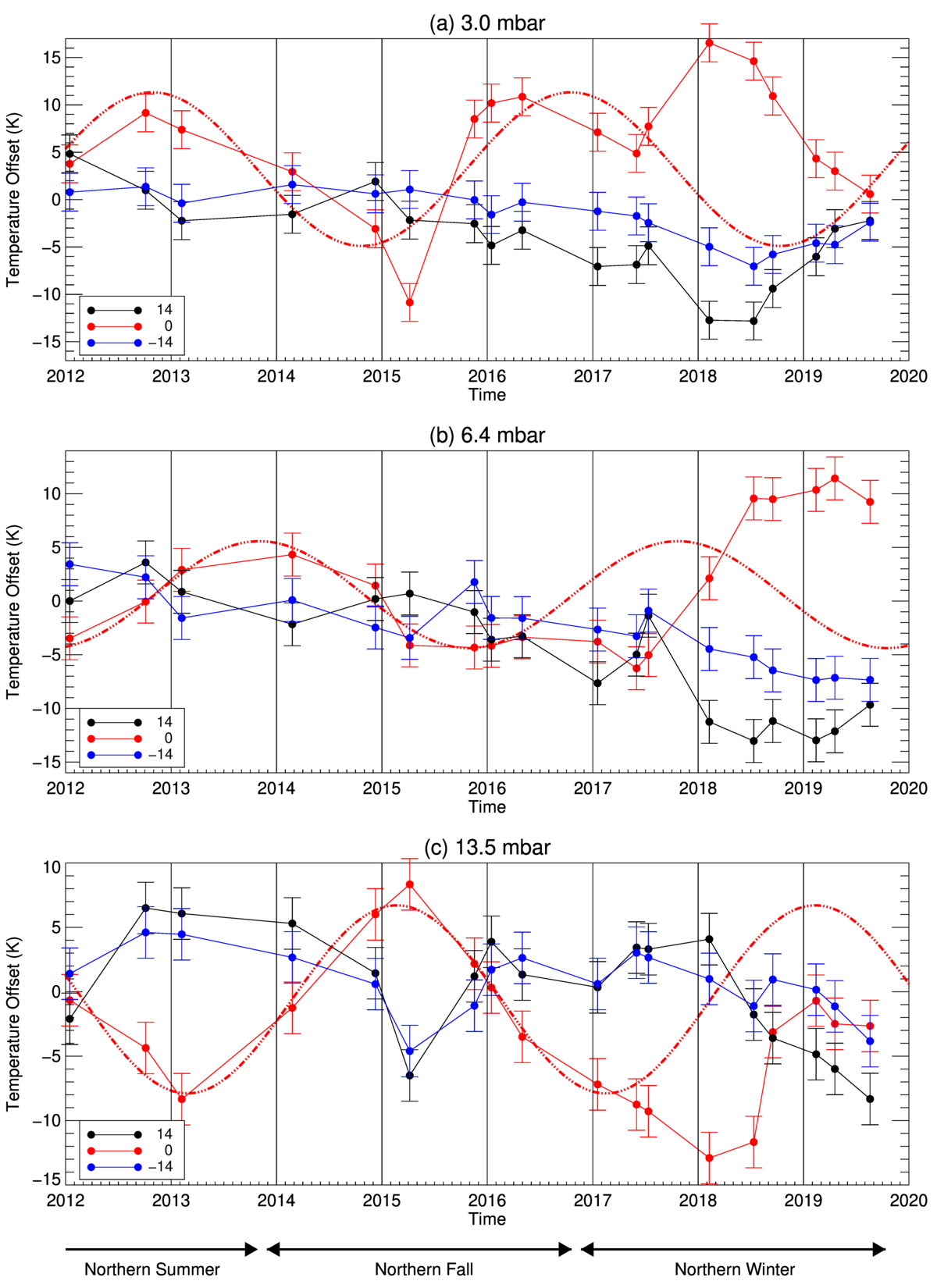}
\caption{Time series of Jupiter's QQO from 2012--2019, showing the temperature offsets at (a) 3.0 mbar, (b) 6.4 mbar and (c) 13.5 mbar. Each plot shows retrieved temperatures from three different latitudes: 14$^{\circ}$N (black), 0$^{\circ}$ (red) and 14$^{\circ}$S (blue). The temperature offset for each pressure level is defined as the zonally-averaged temperature relative to the average of the maximum and minimum temperature during the first four years of data (corresponding to a complete, unperturbed oscillation). The dashed red lines show a sinusoidal curve with a period of 4.0 years (the best-fit period for the 2012--2017 13.5-mbar data). Jupiter's seasons are labelled at the bottom.}
\label{fig:time_series}
\end{figure*}

Figure~\ref{fig:time_series} presents the QQO data for three different pressure levels: 3.0 mbar, 6.4 mbar and 13.5 mbar. The red data points correspond to the same equatorial data that are shown in Figure~\ref{fig:time_series_contour}. The dashed lines in Figure~\ref{fig:time_series_contour} highlight the three pressure levels used in Figure~\ref{fig:time_series}. Alongside the equatorial temperature offsets, we also present the temperature offsets for 14$^{\circ}$N (black) and 14$^{\circ}$S (blue).

The long-term 7.8-\textmu m images that were first used to identify the QQO~\citep{orton91,leovy91} are sensitive to the 10--20 mbar pressure range; Figure~\ref{fig:time_series}(c) therefore provides the best comparison with those results. This pressure levels also corresponds to the region of maximum sensitivity for the TEXES data. Figure~\ref{fig:time_series}(c) shows a clear sinusoidal pattern for the first five years of the data collection (2012--2017). Fitting a sine curve to the pre-2017 red equatorial data results in a period of 4.0$\pm$0.2 years; the fitted curve is shown by the red dashed line. The amplitude of this fitted oscillation at 13.5 mbar is 7$\pm$1 K. The 14$^{\circ}$N and 14$^{\circ}$S temperatures also show a roughly sinusoidal pattern at 13.5 mbar for the first five years, which is in anti-phase to the equatorial temperatures and has a comparable magnitude. This agrees with previous IRTF observations which showed that the 7.8-\textmu m brightness is anti-correlated between the equator and $\pm$13$^{\circ}$~\citep{friedson99}. 

As previously discussed in \citet{cosentino17}, the vertically-resolved TEXES observations show for the first time that Jupiter's QQO extends to lower pressures. This can be clearly seen in Figure~\ref{fig:time_series_contour}, which shows the diagonal, downward propagating pattern extend between $\sim$1 mbar and $\sim$20 mbar. At pressures less than 1 mbar, no clear oscillation is observed, although it should be noted that our sensitivity begins to decrease at these pressures.

Figures~\ref{fig:time_series}(a) and (b) show how the QQO oscillations appear at pressures of 3.0 mbar and 6.4 mbar. The 2012--2017 data in these figures show that the equatorial temperatures at these lower pressures still exhibit an approximately sinusoidal pattern. The dashed red lines in Figures~\ref{fig:time_series}(a) and (b) show the best-fit sine curve to the 2012--2017 equatorial data, assuming the same 4.0-year period derived from Figure~\ref{fig:time_series}(c). At 6.4 mbar, the oscillation is still a smooth sine wave; at 3.0 mbar, it appears to be more `sawtooth'. At these higher altitudes, any oscillation in the off-equatorial temperatures is significantly weaker than at 13.5 mbar.

\subsection{Disruption of the QQO in 2017}
\label{sec:disruption}

The temperature offsets in 2012--2017 appear to show a `typical' QQO pattern, consisting of a smoothly varying sinusoidal oscillation with a period of $\sim$4 years. However, Figures~\ref{fig:time_series_contour} and~\ref{fig:time_series} show that there is a phase-shift disruption to this pattern in mid-2017.

The equatorial data, shown in red in Figure~\ref{fig:time_series}(c), show that the temperatures continue to decrease in late 2017 / early 2018, rather than following the 4-year-period sinusoidal pattern shown by the dashed red line. This leads to the minimum in the equatorial 13.5-mbar temperature occurring $\sim$1 year later than expected. Similar shifts can be observed in Figures~\ref{fig:time_series}(a) and (b). In Figure~\ref{fig:time_series}(b), the equatorial data lags $\sim$1 year behind the nominal sinusoid from 2017 onwards. In Figure~\ref{fig:time_series}(a), the change is even more stark. In 2017--2018, we would expect the equatorial 3-mbar temperatures to be decreasing, reaching a minimum in late 2018. Instead, the temperatures start to decrease as expected, but then suddenly tick upwards in mid-2017, reaching a local maximum in early 2018, just two years after the previous local maximum. When the equatorial temperatures start to decrease again, they are $\sim$1 year behind schedule.

Figure~\ref{fig:time_series} also shows how the off-equatorial temperatures vary from 2017 onward. Figure~\ref{fig:time_series}(c) shows that the $\pm$14$^{\circ}$ temperatures broadly continue in anti-phase with the equatorial temperatures. In particular, it can be seen that the 14$^{\circ}$N temperature peaks in early 2018, at the same time as the equatorial temperatures reach their delayed minimum. The off-equatorial results shown in Figures~\ref{fig:time_series}(b) and (c) also show a change in 2017. Before 2017, the $\pm$14$^{\circ}$ temperature offsets are relatively flat. From mid-2017 onward, the 14$^{\circ}$S temperatures continue to be mostly flat, but the 14$^{\circ}$N temperatures show a stronger inverse correlation with the equatorial temperatures. 

Further insight into how the equatorial temperatures evolved over this time period can be obtained from Figure~\ref{fig:time_series_contour}. The solid white line shows the motion of the descending warm branch of the QQO. In 2017, this warm branch undergoes an abrupt upward displacement. This has the effect of `delaying' the QQO by $\sim$1 year at the 3.0-mbar, 6.4-mbar and 13.5-mbar pressure levels, as shown in Figure~\ref{fig:time_series}. The upwards displacement of the descending warm branch coincides with an abrupt, short-lived warming at high altitudes (0.2--2 mbar).

\begin{figure*}
\centering
\includegraphics[width=16cm]{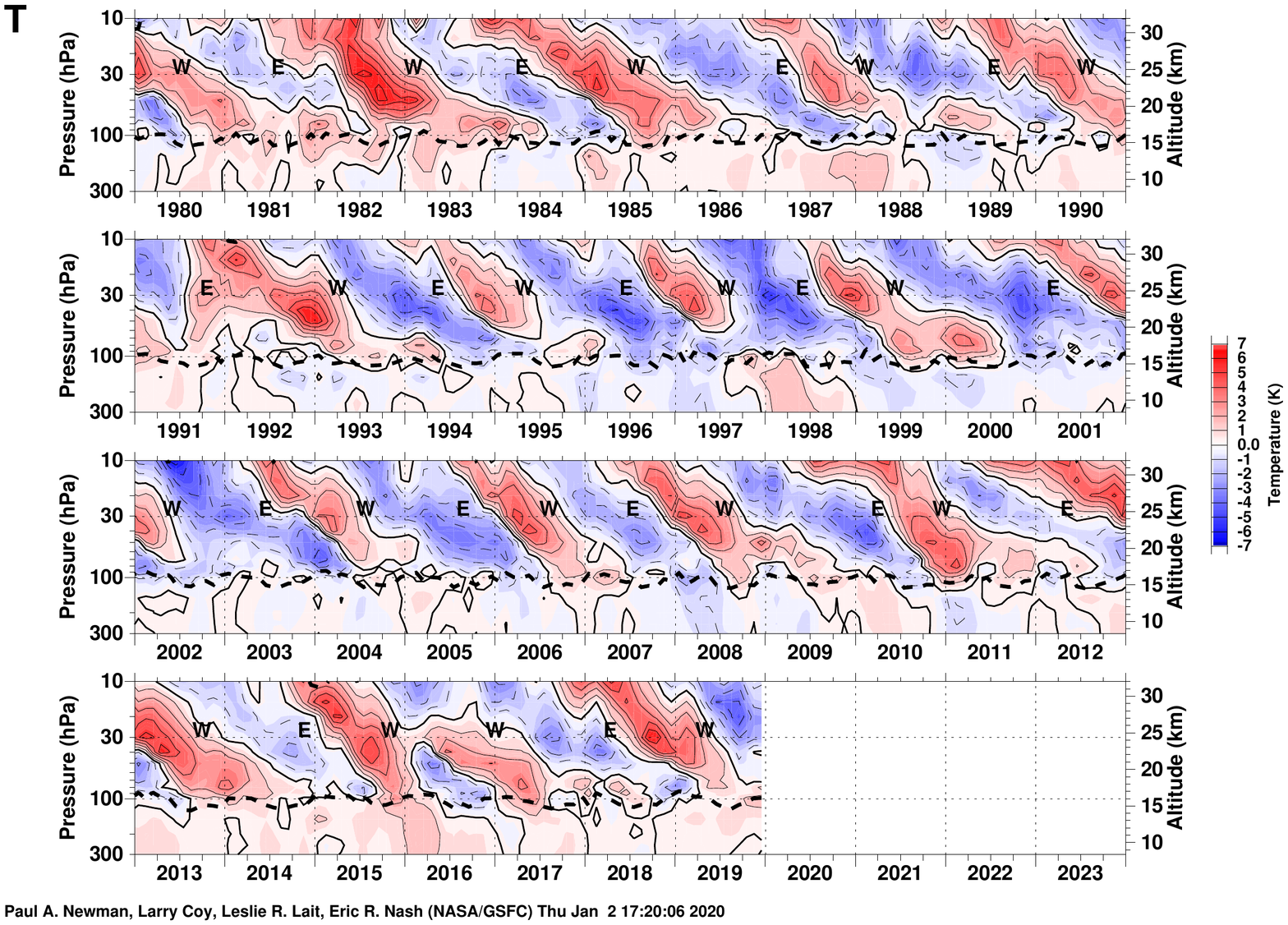}
\caption{Earth's equatorial QBO from 1980--2020; figure from NASA/GSFC Data Services (https://acd-ext.gsfc.nasa.gov/Data\_services/met/qbo/qbo.html). Radiosonde temperature data were obtained from the Meteorological Service Singapore Upper Air Observatory and the data have been detrended and the annual cycle has been removed. The 2015--2016 QBO disruption can be clearly seen~\citep{newman16}.}
\label{fig:qbo}
\end{figure*}

The QQO disruption observed in Figures~\ref{fig:time_series_contour} and~\ref{fig:time_series} bears some similarities to the disruption of the Earth's QBO that was observed in 2015--2016~\citep{osprey16,newman16}. The QBO has been observed continuously using tropical radiosonde wind observations since 1953 and is ordinarily one of the most repeatable phenomena in the Earth's atmosphere, with an oscillation period that has varied between 22 and 34 months~\citep{baldwin01}. In 2015--2016, a disruption was observed in the typical QBO pattern for the first time in over sixty years of observations. \citet{osprey16} and \citet{newman16} show the QBO disruption in terms of the zonal winds; Figure~\ref{fig:qbo} shows the disruption in terms of the stratospheric temperature anomalies, allowing it to be directly compared with Figure~\ref{fig:time_series_contour}. \citet{newman16} describe the disruption as an apparent upward displacement of an anomalous westerly wind and the development of easterlies at 40 mbar; this has the effect of altering the short-term behavior of the oscillation, producing an unusually long-lasting 23-month eastward phase at 20 mbar, compared to the average 14-month duration~\citep{kumar18}. In the temperature data shown in Figure~\ref{fig:qbo}, this manifests as an upward displacement in the warm branch and the development of a new cold branch at 40 mbar.

This is morphologically similar to Figure~\ref{fig:time_series_contour}, which shows an upward displacement of the QQO's descending warm branch in 2017. As with the 2015--2016 QBO disruption, this has the effect of altering the short-term behavior of the QQO, producing an unusually long cold phase at 13.5 mbar. However, we should note that the QBO disruption in 2015--2016 was a highly unusual event; the earth's wind patterns and temperatures have been consistently tracked for over 60 years, providing a vertically-resolved long-term dataset, and Figure~\ref{fig:qbo} shows that this was a unique event within that time period. In contrast, it is unclear whether the QQO disruption shown in Figure~\ref{fig:time_series_contour} is an uncommon event, as vertically-resolved measurements of the QQO have only been made since 2012. 7.8-\textmu m imaging data have been made since 1980~\citep{orton91,leovy91}, but these observations do not provide vertical resolution and are therefore not able to fully capture the effect of a vertical displacement of the QQO; it is possible that some of the variability in the oscillation period observed in the imaging data~\citep{simon-miller06} is due to a similar type of disruption. 

\subsection{Potential causes of the 2017 disruption}
\label{sec:anomaly}

Based on Figures~\ref{fig:time_series_contour} and~\ref{fig:time_series}, we conclude that there was a disruption to the QQO pattern in 2017; this disruption consisted of an upward displacement of the descending warm branch, leading to a phase shift in the oscillation. Numerical modeling of the observed disruption is beyond the scope of this paper and will be a subject for future study. Instead, we discuss here the causes of similar disruptions on other planets and present observations of a high-altitude thermal anomaly that may be linked to the observed disruption.

\citet{barton17} found that the Earth's QBO disruption in 2015--2016 was caused by the propagation of Rossby waves from the extratropical Northern Hemisphere. The combination of the timing of the QBO relative to the annual cycle and an extreme El Ni\~no event meant that the subtropical easterly jet in the winter lower stratosphere was unusually weak; this allowed westward-moving Rossby waves to propagate southwards into the equatorial region, and deposit a large flux of westward momentum at 40 mbar.

Like the Earth and Jupiter, Saturn also has an oscillation in the equatorial stratosphere. This oscillation is known as Saturn's Semi-Annual Oscillation (SSAO) and it has a period of 14.7$\pm$0.9 years~\citep{orton08,fouchet08}. \citet{fletcher17b} found that in 2011--2014, the SSAO was disrupted by an intense storm in the northern mid-latitudes. The storm consisted of powerful convective plumes in the troposphere, and led to the formation of a large, hot, westward-moving vortex in the stratosphere at 40$^{\circ}$N~\citep{fletcher12}. This vortex (known as the `beacon') persisted for over three years, significantly longer than the tropospheric storm. At its peak, the beacon had a temperature that was 80 K higher than the surrounding quiescent regions at 2 mbar. During the same time period, \citet{fletcher17b} showed that the SSAO signature was essentially absent, before resuming in 2014. As with the Earth's QBO disruption,~\citet{fletcher17b} attribute the SSAO disruption to an injection of westward momentum in the equatorial regions by extratropical waves; in this case, the waves were thought to be driven by the hot beacon and/or the underlying storm. 

As both the QBO disruption and the SSAO disruption were caused by the propagation of extratropical waves, we used the TEXES zonally-resolved temperatures maps to search for any unusual behavior that could be associated with the 2017 QQO disruption. We found a high-altitude, mid-latitude thermal anomaly in the May/June 2017 data, the same observing run in which the disruption was first observed; an inspection of the spatially-resolved temperature maps from the other observing runs showed that this was the largest thermal anomaly observed over the eight year time period.

\begin{figure*}
\centering
\includegraphics[width=15cm]{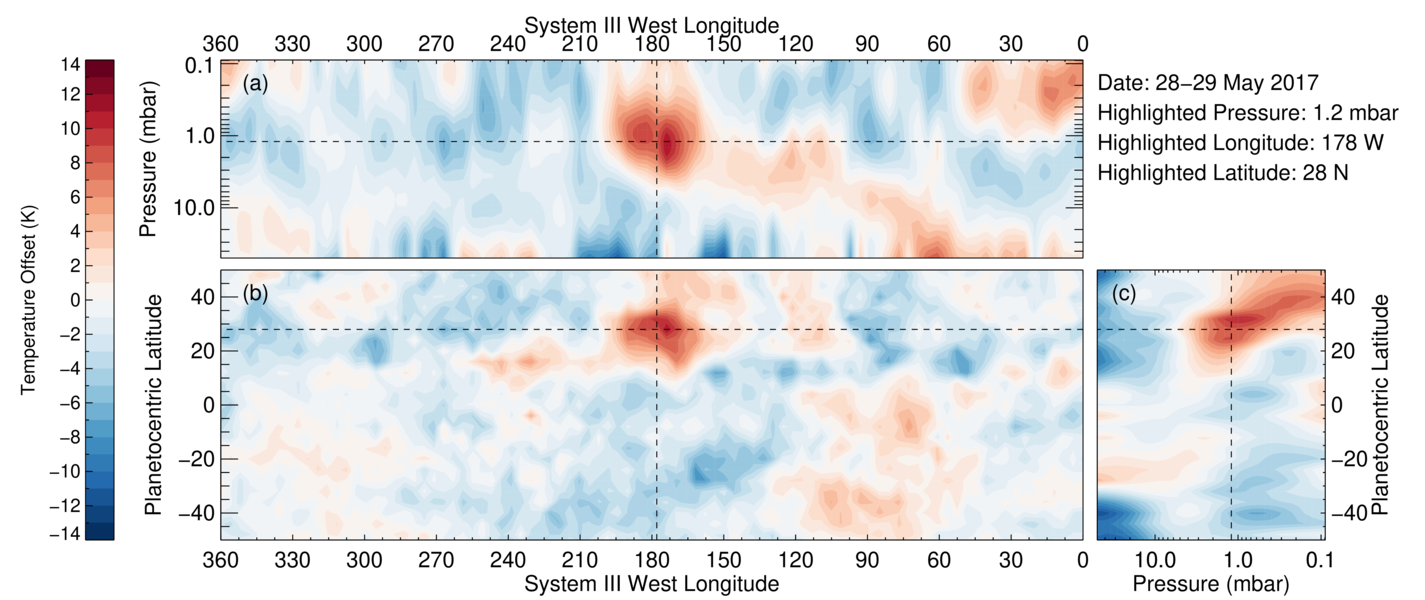}
\caption{A warm thermal anomaly, as seen on May 28--29 2017. The three frames show three orthogonal slices through a three-dimensional retrieved temperature map; these slices are centered on a warm thermal anomaly. The temperature offset is defined as the temperature relative to the mean zonal temperature at a given pressure and latitude level.}
\label{fig:anomaly_may28}
\end{figure*}

\begin{figure*}
\centering
\includegraphics[width=15cm]{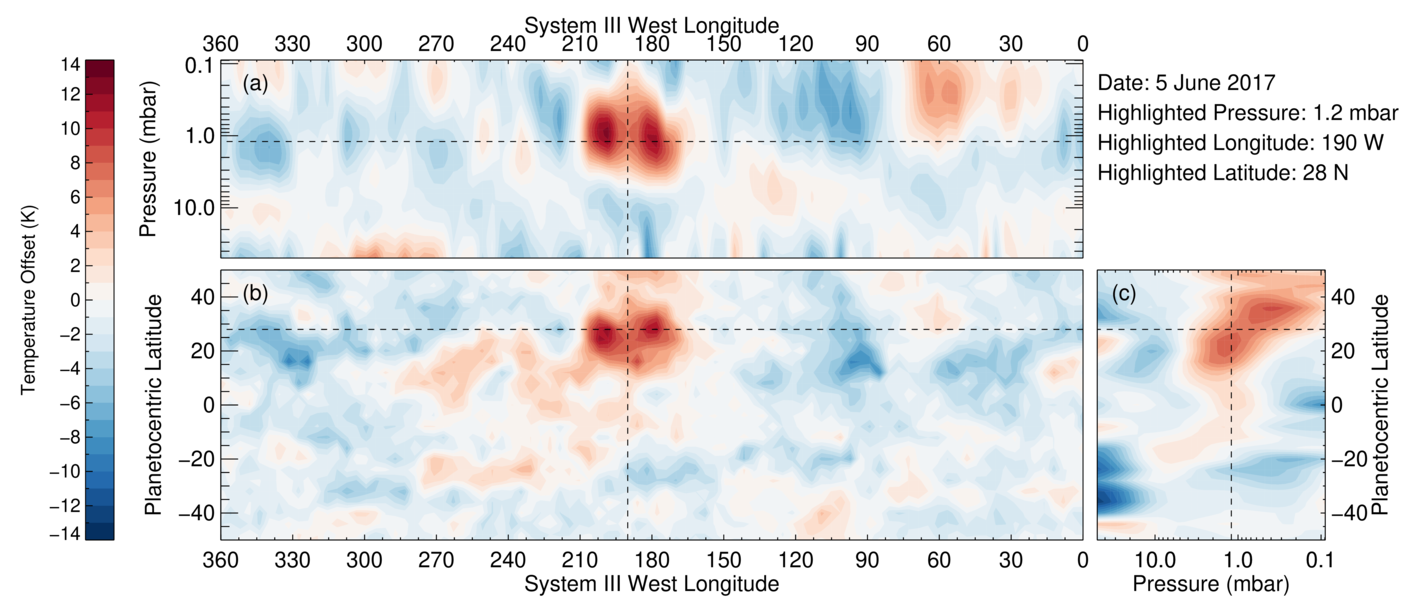}
\caption{The same thermal anomaly shown in Figure~\ref{fig:anomaly_may28}, as seen on June 5 2017.}
\label{fig:anomaly_june5}
\end{figure*}

The May/June 2017 thermal anomaly is shown in Figures~\ref{fig:anomaly_may28} and~\ref{fig:anomaly_june5}. Figure~\ref{fig:anomaly_may28} shows three different slices through the three-dimensional temperature map obtained using data from May 28--29. The temperature offset is defined as the temperature relative to the mean zonal temperature at a given pressure and latitude level. The average retrieved error on the temperature offsets is $\sim$2 K. Figure~\ref{fig:anomaly_may28}(a) shows the thermal anomaly as a function of pressure and longitude, at a latitude of 28$^{\circ}$N; ~\ref{fig:anomaly_may28}(b) shows the thermal anomaly as a function of latitude and longitude, at a pressure of 1.2 mbar; and ~\ref{fig:anomaly_may28}(c) shows the thermal anomaly as a function of latitude and pressure, at a longitude of 178$^{\circ}$W. In each case, the dashed black lines highlight the latitude, longitude and pressure of interest; the three images intersect at the intersection of the dashed black lines. Figure~\ref{fig:anomaly_june5} shows a comparable set of images for data obtained on June 5. 

Both Figures~\ref{fig:anomaly_may28} and~\ref{fig:anomaly_june5} show a warm high-altitude double-peaked thermal anomaly centered at a latitude of 28$^{\circ}$N and a pressure of 1.2 mbar. This is located in Jupiter's North Temperate Belt, northwards of the planet's strong prograde jet at 23$^{\circ}$N~\citep{porco03}. By fitting a Gaussian curve to the retrieved temperatures as a function of longitude, we find that the thermal anomaly moves westward with a velocity of 1.5$\pm$0.3 degrees/day; at a latitude of 28$^{\circ}$N, this is equivalent to 19$\pm$4 ms\textsuperscript{-1}. No anomaly was present at the same latitude and pressure region during the January 2017 or February 2018 observing runs. It was also not observed in July 2017, although full longitudinal coverage was not obtained on this occasion and the relevant longitudes were not observed (assuming a constant velocity of 1.5 degrees/day). 

The correlation between the timing of this thermal anomaly and the disruption of the QQO is suggestive of a link between these two events. In addition, the thermal anomaly has some features in common with the stratospheric `beacon' on Saturn that disrupted the SSAO; this thermal anomaly is also a westward-moving hotspot in the stratospheric mid-latitudes. However, the scales of the two hotspots are very different. Saturn's beacon was 80 K warmer than its surroundings and persisted for over three years, while the thermal anomaly we observe here was 13 K warmer than its surroundings and persisted for less than seven months. It is unclear whether a thermal anomaly on this smaller scale would have the capacity to influence the equatorial temperatures.

Jupiter's troposphere undergoes large-scale upheavals relatively frequently. These events have the ability to dramatically change the appearance of the planet by altering the cloud structure~\citep{sanchez-lavega08,fletcher17c} and would be plausible candidates for driving a disruption in the QQO and/or causing localized thermal anomalies in the stratosphere; rather than causing the disruption directly, one possibility is that both the thermal anomaly and the QQO disruption are the result of a localized storm deep in Jupiter's troposphere. However, there were no notable planetary scale disturbances during the 2017--2018 timeframe that we can easily tie to the QQO disturbance, despite an extensive search of images from different sources.

\citet{sanchez-lavega17b} report on a large disturbance in Jupiter's North Temperate Belt in October 2016, consisting of a set of four powerful plumes located at 23$^{\circ}$N, that moved eastward with time. However, this activity had ceased by late November 2016, seven months before the QQO disturbance and the mid-latitude thermal anomaly were observed.
\citet{wong20} present an archive of high spatial resolution images of Jupiter in the 2016--2020 time period; their observations provide good temporal coverage in the spring--summer 2017 timeframe and include full longitudinal coverage in the visible from Hubble's Wide Field Camera 3 in April 2017, but they note no large convective storm in the northern low or mid latitudes beside the one observed in October 2016. \citet{antunano18} describe a pattern of semi-regular disturbances in Jupiter's equatorial zone, occurring with a periodicity of 6--8 or 13--14 years. These disturbances consist of a dramatic clearing in Jupiter's equatorial cloud layer, resulting in a sharp increase in the 5-\textmu m brightness. The most recent disturbance described in~\citet{antunano18} occurred in 2006--2007, and they predicted a new equatorial zone disturbance in 2019--2021. Changes in the 5-\textmu m brightness began to be observed in late 2018~\citep{orton19}, over a year after the QQO disruption. Finally, we consulted the archives of the British Astronomical Association (BAA\footnote{britastro.org}), who keep a detailed record of observations of Jupiter made by amateur astronomers. BAA reports carefully document the outbreak in the North Temperate Belt in October 2016, but there was no similar outbreak observed in spring/summer 2017, despite a large dataset of observations~\citep{rogers17}. 
 
While Jupiter's troposphere exhibits constant activity, the 2017 QQO disruption does not appear to coincide with any large global-scale tropospheric upheaval events that could be seen in visible or infrared observations. It is possible that a storm could have been present deeper in Jupiter's atmosphere, below the visible cloud deck, but it is difficult to see how such a storm could affect stratospheric temperatures without causing a visible plume. Further monitoring of Jupiter's stratospheric and tropospheric temperatures will be required to make further progress in understanding the causality of such events in Jupiter's atmosphere.


\section{Conclusions}
\label{sec:conclusions}

In this paper, we used high-resolution mid-infrared data from TEXES/IRTF to track the evolution of Jupiter's quasi-quadrennial oscillation (QQO) over the course of approximately two full cycles (2012--2019). We used a radiative transfer model to retrieve the zonally-averaged and zonally-resolved atmospheric temperatures in the 0.01--30 mbar pressure range. Using the temperature maps obtained from these retrievals, we come to the following conclusions:

\begin{enumerate}

    \item Between 2012 and 2017, the 13.5-mbar equatorial temperatures had a smoothly sinusoidal pattern with a period of 4.0$\pm$0.2 years and an amplitude of 7$\pm$1 K at 13.5 mbar (our region of maximum sensitivity). During this same time period, the 13.5-mbar temperatures at $\pm$14$^{\circ}$ were in anti-phase with the equatorial temperatures. This is consistent with previous studies of the QQO~\citep{orton91, friedson99}.
    
    \item Between 2012 and 2017, the 6.4-mbar and 3.0-mbar equatorial temperatures also displayed an approximately sinuoidal pattern, with a similar period. This extension of the QQO up to these lower pressure levels was a new result presented in~\citet{cosentino17} using the 2012--2016 TEXES data.
    
    \item In 2017, there was a disruption to the QQO pattern. This disruption is evident at 3-20 mbar, and appears to manifest as an abrupt upward displacement of the descending warm branch. This disruption has similarities to the disruption of the Earth's QBO in 2015--2016.
    
    \item In May/June 2017, we observed a localized warm thermal anomaly located at a latitude of 28$^{\circ}$N and a pressure of 1.2 mbar. By comparing observations made $\sim$1 week apart, we found that this thermal anomaly was moving westward with a velocity of 19$\pm$4 ms\textsuperscript{-1}
    
\end{enumerate}
    
Based on the timing of the warm thermal anomaly, there is a possibility that it may be related to the QQO disruption that we observe. We note that a hot thermal anomaly in Saturn's mid-latitude stratosphere was linked to a disruption in the SSAO. However, the warm anomaly that we observe in Jupiter's stratosphere is significantly cooler ($\sim$13 K) and it is unclear whether it would have a significant impact on the equatorial regions. One possibility is that both the thermal anomaly and the QQO disruption are the result of storms deep in Jupiter's troposphere, although no unusual activity was observed during this time period. Future modelling work is required to better understand the factors that drive changes to Jupiter's QQO.


\section*{Acknowledgments}

This research was primarily funded by NASA Planetary Astronomy (PAST) grant NNX14AG34G. GSO was supported by funds from NASA distributed to the Jet Propulsion Laboratory, California Institute of Technology. This work was based on data obtained with the NASA Infrared Telescope Facility, which is operated by the University of Hawaii under a contract with the National Aeronautics and Space Administration. The authors would like to acknowledge the contributions of many observing collaborators and NASA/IRTF staff. The authors would also like to thank the two anonymous reviewers whose comments helped improve and clarify this manuscript.



\end{document}